\documentclass[pra,twocolumn,amsmath,amssymb,superscriptaddress]{revtex4}

\def\Per{\mathop{\rm Per}}

\begin{document}

\title{Polynomial computational complexity of matrix elements of finite-rank-generated single-particle operators
in products of finite bosonic states}

\author{Dmitri~A.~Ivanov}
\affiliation{Institute for Theoretical Physics, ETH Z\"urich,
8093 Z\"urich, Switzerland}

\begin{abstract}
It is known that computing the permanent of the matrix $1+A$, where $A$ is a finite-rank matrix, requires
a number of operations polynomial in the matrix size. Motivated by the boson-sampling proposal of restricted quantum
computation, I extend this result to a generalization of the matrix permanent: an expectation value in a product
of a large number of identical bosonic states with a bounded number of bosons. This result complements
earlier studies on the computational complexity in boson sampling and related setups. The proposed
technique based on the Gaussian averaging is equally applicable to bosonic and fermionic systems. This also allows
us to improve an earlier polynomial complexity estimate for the fermionic version of the same problem.
\end{abstract}

\date{May 29, 2023}

\maketitle

\section{Introduction}

Potential implications for quantum computing motivated an interest in computational complexity
of quantum-mechanical amplitudes in many-particle systems.
The best known example is the boson-sampling proposal in Refs.\ \onlinecite{aaronson-2011,aaronson-2013}:
since the multi-particle amplitudes for non-interacting bosons are given by permanents, which are presumably
hard to compute \cite{valiant-1979}, it was conjectured that
boson-sampling experiments are hard to model on classical computers (the connection between the
two complexities is not straightforward though, since sampling is not equivalent to computing
probabilities).

The boson-sampling proposal raised two interesting questions of general relevance to quantum information
theory. The first one was the origin of computational complexity in apparently simple non-interacting
quantum mechanics. The second one was an apparent asymmetry between bosons and fermions (for which
a similar setup produced an easily computable determinant).

Those questions were addressed and mainly answered in Refs.\ \onlinecite{ivanov-2017,ivanov-2020}. 
There, we extended the boson-sampling construction to a wider class of multi-particle product states,
either bosonic or fermionic. We further analyzed the computational complexity of matrix elements
of a general non-interacting evolution in such product states. The examples considered
in those works suggest that the \#P complexity class of computing
a matrix permanent may also be the general complexity class for this problem, except in very special
cases (e.g., for Gaussian states). In other words, the source of computational complexity is
the quantum (non-Gaussian) nature of the underlying states. In this respect, bosons and fermions
behave similarly in terms of computational complexity.

It was also shown in Ref.~\onlinecite{ivanov-2020} that, for generating computational complexity,
the noninteracting evolution must be sufficiently complex, involving a large number of degrees
of freedom. In contrast, a finite-rank evolution (defined as evolution
operators of the form $1+A$, where $A$ is a finite-rank matrix)
does not generate amplitudes of high computational complexity. Instead, such operators only have
matrix elements (between multi-particle product states) computable in polynomial time. As a particular case,
the lemma about the polynomial computability of the permanent $\Per(1+A)$ with a finite-rank matrix $A$
was proven \cite{aaronson-2013,ivanov-2020}.

The results of Ref.~\onlinecite{ivanov-2020} on the finite-rank evolution missed however one case:
that of a general bosonic product state.
In this paper, I rectify this omission and prove the polynomial computability of the
expectation value of the non-interacting multiplicative extension of the operator $1+A$ with a finite-rank matrix $A$
in any ``finite'' bosonic product state (a ``finite'' state means here that each of the factors has
a bounded number of particles). The number of operations required for such a computation is estimated as
$O(N^{2k+1})$, where $k$ is the rank of $A$. This estimate coincides with the particular case of the permanent
proven in Ref.~\onlinecite{ivanov-2020}.

The proof generalizes that in Ref.~\onlinecite{ivanov-2020}
by explicitly introducing Gaussian averaging of auxiliary polynomials.
This way it can also be viewed as a generalization of the ``permanent as an inner product'' construction
of Refs.~\onlinecite{marcus-1961,marcus-1962}.

The proof may be performed in parallel for bosons and fermions. As a byproduct,
it also allows us to improve the polynomial estimate for the fermionic case. In Ref.~\onlinecite{ivanov-2020},
the calculation of the finite-rank-evolution matrix elements was estimated to require $O(N^{2k})$ operations,
and now we replace this estimate by $O(N)$ (independent of the rank $k$, assuming $k$ being a finite number which
does not grow with $N$).

The paper is organized as follows. In Section \ref{sec:construction-results}, I formulate and prove
the main results. In Section \ref{sec:examples}, I comment on the application
of the method to the simplest examples: the matrix permanent in the bosonic case
and the determinant in the fermionic case. Finally, Section \ref{sec:conclusion} contains concluding
remarks and comments.

\section{Construction and results}
\label{sec:construction-results}

\subsection{Definition: a multiplicative extension of a single-particle operator}

Let $U$ be a single-particle operator. We denote by $P(U)$ its {\em multiplicative extension} in the
bosonic (or fermionic) Fock space. It is defined by its action on the standard basis in the Fock space:
\begin{equation}
P(U) a^\dagger_{j_1}\ldots a^\dagger_{j_k}|\star\rangle =
\sum_{i_1\ldots i_k} U_{i_1,j_1} \ldots U_{i_k,j_k} a^\dagger_{i_1}\ldots a^\dagger_{i_k} |\star\rangle\, ,
\label{PU-def}
\end{equation}
where $|\star\rangle$ is the vacuum (zero-particle) state and $a^\dagger_i$ are the bosonic (or fermionic, respectively)
creation operators.

\subsection{Multiplicative extension as an evolution operator}

One can easily prove the following property: if $U$ is invertible, then
\begin{equation}
P(U)=\exp(\sum_{ij} a^\dagger_i (\ln U)_{ij} a_j)\, ,
\label{PU-exponent}
\end{equation}
This relation is not used in our contruction below, but it
explains the physical meaning of the multiplicative extension $P(U)$: if $U$ is a single-particle
evolution operator, then $P(U)$ is the corresponding noninteracting evolution of the multi-particle states.
It is valid equally for bosons and fermions and can be proven by directly commuting the right-hand side
of Eq.~(\ref{PU-exponent}) with the creation operators in the definition (\ref{PU-def}).

\subsection{Lemma: multiplicative extension as a normal-ordered exponent}

\begin{equation}
P(1+A)= : \exp(\sum_{ij} a^\dagger_i A_{ij} a_j) : \, ,
\label{lemma-2}
\end{equation}
where $::$ is the normal ordering: all the creation operators are on the left and all the annihilation
operators on the right. This relation is also equally valid for bosons and fermions and can be proven by directly
verifying the definition (\ref{PU-def}).

\subsection{Definition: Gaussian averaging of a polynomial}

For any polynomial $f(z_1,\ldots, z_k, z_1^*, \ldots z_k^*)$ define its {\em Gaussian average} as
\begin{multline}
S(f)=\int \left( \prod_{i=1}^k e^{-z_i z_i^*} dz_i\; dz_i^* \right) \\
f(z_1,\ldots, z_k, z_1^*, \ldots z_k^*)\, ,
\end{multline}
where the integration measure $dz_i\; dz_i^*$ over the complex conjugate variables is normalized so that
\begin{equation}
\int e^{-z z^*} dz\; dz^*\; z^m (z^*)^n = \delta_{m,n} m!\, .
\label{gaussian-norm}
\end{equation}
This construction will be used in relation to the bosonic case below.
For the fermionic case, exactly the same construction will be applied with the independent
Grassmann (anticommuting) variables $z_i$ and $z_i^*$ assuming the standard rules of the Grassmann-variable
calculus (with $n$ and $m$ limited to be either 0 or 1) \cite{berezin-1966}.

\subsection{Lemma: Finite-rank evolution as a Gaussian average}

Let $A$ be a matrix of rank $k$,
\begin{equation}
A_{ij}=\sum_{\alpha=1}^k u_{i\alpha} v_{\alpha j}\, .
\label{uv-def}
\end{equation}
Then
\begin{equation}
P(1+A)=S\left(
e^{\sum_{\alpha,i}  z_\alpha u_{i\alpha} a^\dagger_i}
e^{\sum_{\alpha,j}  z^*_\alpha v_{\alpha j} a_j}
\right)\, .
\label{lemma-3}
\end{equation}
It can be proven either by comparing the expansions of Eqs.\ (\ref{lemma-3}) and (\ref{lemma-2}) term by term
or, more elegantly, by decoupling the product $uv$ in the exponent of the right-hand side of Eq.~(\ref{lemma-2})
by Gaussian integration (known to physicists as Hubbard-Stratonovich transformation).
This statement, similarly to the previous lemmas, holds for both bosonic and fermionic cases, with the only
difference that the variables $z_i$ and $z_i^*$ are commuting in the bosonic case and anticommuting (Grassmann)
in the fermionic case.

\subsection{Definition: finite bosonic and fermionic states}

We call a bosonic or fermionic state {\em finite}, if (A) it is a state in a Fock space built on a finite
number of single-particle states and (B) in the bosonic case, the total number of particles in this state
is bounded. A finite state has only a finite number of nonzero components in the standard basis
[used, e.g., in Eq.(\ref{PU-def})]. The condition (B) only makes sense for bosonic states: any
fermionic state built on a finite-number of single-particle states automatically has a bounded number
of particles because of the Pauli principle (not more than one particle per single-particle state).

\subsection{Lemma: Expectation value of a finite-rank evolution matrix in a product state
via Gaussian averaging of auxiliary polynomials}

Consider a $d$-dimensional single-particle space $L_1$ ($d$ is finite). Let us denote the corresponding bosonic or
fermionic Fock (multi-particle) space $F_1$. Let $\Psi_1 \in F_1$ be a finite state (it is automatically
finite in the fermionic case). Now take $N$ identical copies
of $L_1$. Then we can consider the $(Nd)$-dimensional single-particle space
\begin{equation}
L=L_1 \oplus \ldots \oplus L_N\, ,
\label{L-product}
\end{equation}
where each $L_\mu$ in the sum is identical to $L_1$. The corresponding Fock space is
the product
\begin{equation}
F=F_1 \otimes \ldots \otimes F_N
\label{Fock-product}
\end{equation}
of $N$ identical factors.
The state $\Psi_1$ may also be replicated into the product state in $F$,
\begin{equation}
\Psi=\Psi_1 \otimes \ldots \otimes \Psi_N\, ,
\label{Psi-product}
\end{equation}
again with $N$ identical states in the Fock spaces $F_\mu$.

Let $A$ be a square matrix of size $Nd$ of finite rank $k$ and let $u_{i\alpha}$ and $v_{\alpha j}$ be
defined via its decomposition (\ref{uv-def}). Introduce the decomposition of the corresponding creation
and annihilation operators in the spaces $L_\mu$,
\begin{align}
\hat{u}^\dagger_{\alpha} &= \sum_{i=1}^{Nd} u_{i\alpha} a^\dagger_i = \sum_{\mu=1}^N \hat{u}^\dagger_{\mu\alpha} \, ,
\nonumber\\
\hat{v}_{\alpha} &= \sum_{j=1}^{Nd} v_{\alpha j} a_j = \sum_{\mu=1}^N \hat{v}_{\mu\alpha} \, ,
\end{align}
which defines operators $\hat{u}^\dagger_{\mu\alpha}$ and $\hat{v}_{\mu\alpha}$ acting in the corresponding
spaces $L_\mu$.

Then we have the following result for the expectation value
\begin{equation}
\left\langle \Psi \right| P(1+A) \left| \Psi \right\rangle = 
S\left( \prod_{\mu=1}^N f_\mu(z_1,\ldots, z_k, z_1^*, \ldots z_k^*) \right) \, ,
\label{result-S}
\end{equation}
where
\begin{multline}
f_\mu(z_1,\ldots, z_k, z_1^*, \ldots z_k^*)= \\
\left\langle \Psi_\mu \right|
\exp\left(\sum_{\alpha=1}^k \hat{u}^\dagger_{\mu\alpha} z_\alpha \right)
\exp\left(\sum_{\alpha=1}^k \hat{v}_{\mu\alpha} z_\alpha^* \right)
\left| \Psi_\mu \right\rangle
\label{result-F}
\end{multline}
As in the previous Lemma, the variables $z_i$ and $z_i^*$ are taken to be commuting the bosonic case
and anticommuting (Grassmann) in the fermionic case.

Proof: a direct consequence of the previous Lemma.

\subsection{Main result: Polynomial complexity for finite-rank evolution, bosonic case}
\label{subsec:bosonic-main}

The expectation value (\ref{result-S}) in the above Lemma, in the bosonic case,
is computable in $O(N^{2k+1})$ elementary operations.

Proof: \nopagebreak

(1) The number of particles in each of the states $\left| \Psi_\mu \right\rangle$ is bounded by a small
number independent of $N$, therefore Eq.~(\ref{result-F}) gives a polynomial of small degree for each $\mu$
(only a small number of terms in the expansion of the exponents need to be taken into account). This polynomial
is computable in a finite number of operations for each $\mu$, therefore the total number of operations for
computing all the polynomials (\ref{result-F}) is $O(N)$ and may be neglected in the total complexity estimate.

(2) Computing the expectation value (\ref{result-S}) involves multiplying all the polynomials (\ref{result-F})
and taking the ``diagonal'' coefficients of the result [with matching degrees of $z_\alpha$ and $z_\alpha^*$,
according to Eq.~(\ref{gaussian-norm})]. The degree of the product polynomial in Eq.~(\ref{result-S}) is bounded
by the total number of particles and is therefore $O(N)$. The number of the coefficients of such a polynomial
is $O(N^{2k})$. For each of the $N$ multiplications, we need to update each of those coefficients, which gives
the total estimate on the number of operations $O(N^{2k+1})$.

\subsection{Polynomial complexity for finite-rank evolution, fermionic case}

The expectation value (\ref{result-S}) in the above Lemma, in the fermionic case,
is computable in $O(N)$ elementary operations.

The proof proceeds exactly as in the bosonic case, with the only difference that the degree of the product
polynomial in Eq.~(\ref{result-S}) is bounded by $2k$, because it can be at most of degree one in each of the
anticommuting variables $z_\alpha$ and $z_\alpha^*$. This number does not grow with $N$ and therefore 
scales as $O(1)$. Thus the calculation of the product of $N$ terms in Eq.~(\ref{result-S}) can be done in
$O(N)$ operations.

\section{Examples: permanent and determinant}
\label{sec:examples}

The simplest cases where our construction is applicable are permanent (for bosons)
and determinant (for fermions).

\subsection{Bosonic case: permanent}

In Ref.~\onlinecite{ivanov-2020}, we considered the single-boson state (with $d=1$)
\begin{equation}
|\Psi_1\rangle = a^\dagger |\star\rangle\, .
\end{equation}
The corresponding expectation value is the permanent,
\begin{equation}
\langle \Psi | P(1+A) | \Psi \rangle =\Per(1+A)\, .
\end{equation}
In this case, Eq.~(\ref{result-F}) gives
\begin{multline}
f_\mu(z_1,\ldots, z_k, z_1^*, \ldots z_k^*)= \\
1+
\left(\sum_{\alpha=1}^k u_{\mu\alpha} z_\alpha \right)
\left(\sum_{\alpha=1}^k v_{\alpha\mu} z_\alpha^* \right)\, ,
\label{f-permanent}
\end{multline}
which reproduces the formula in Ref.~\onlinecite{ivanov-2020}.

\subsection{Fermionic case: determinant}

In the case of determinant (the simplest example in the fermionic case), our complexity estimate is consistent
with the so-called Sylvester's determinant theorem:
\begin{equation}
\det (1+UV)= \det (1+VU)\, .
\end{equation}
While the left-hand side is a determinant of size $N$, the right-hand side only has size $k$, the rank of the
matrix $A=UV$. Therefore, the calculation only requires $O(N)$ operations to compute the matrix in the right-hand
side (assuming $k$ is small and does not grow with $N$).

\section{Conclusion}
\label{sec:conclusion}

To summarize, this paper proves the polynomial complexity of the expectation value
$\langle \Psi | P(1+A) | \Psi \rangle$ for a finite-rank matrix $A$ and a finite product state $|\Psi\rangle$.
It completes and elucidates the analysis of Ref.~\onlinecite{ivanov-2020}, further illustrating the
origin of computational complexity in quantum systems.

The reasoning of the bosonic theorem from Section~\ref{subsec:bosonic-main}
fully parallels that of Ref.~\onlinecite{ivanov-2020} for the case of the
matrix permanent and results in the same complexity estimate $O(N^{2k+1})$ for a wider class of bosonic states.
In Ref.~\onlinecite{ivanov-2020}, the Gaussian-average construction was not spelled out,
but could only be guessed from the formulas. Now we could use it explicitly and extend it
to a wider class of states. It can also be viewed as a generalization of the
inner-product formula for the permanent of a matrix product known from Refs.~\onlinecite{marcus-1961,marcus-1962}:
if we only take the higher-order terms in $u$ and $v$ from Eq.~(\ref{f-permanent}), then we reproduce the
inner-product formulas from those works.

In the fermionic case, the new calculation algorithm gives a significant speed up as compared to
Ref.~\onlinecite{ivanov-2020}:
$O(N)$ versus $O(N^{2k})$ operations, which also marks a difference between the bosonic and fermionic calculations.
This difference can, however, be easily understood on physical grounds. Calculating the matrix elements in our
finite-rank problem involves keeping track of multi-particle states in the basis defined by the coefficients
$u$ and $v$. For fermions, it amounts to manipulating $k$ fermionic levels, which gives rise to $2^k$ multi-particle
states (which is $O(1)$ in terms of growing with $N$). For bosons, the the same $k$ single-particle levels may be
occupied with different numbers of particles, and the number of states grows as $N^k$, hence one needs a larger
number of operations to keep track of those states. Nevertheless, we can still speak of a certain symmetry between
bosonic and fermionic problems, since both of them lead to a polynomial computational complexity.

Finally, we considered identical factor states in Eq.~(\ref{Psi-product}) for simplicity only.
The results of the paper trivially extend to the case of product states with different factors
in Eqs.\ (\ref{L-product})--(\ref{Psi-product}), provided the number of particles in each of them
is bounded uniformly and independently of $N$. An extension to the case of different bra- and ket-states
(to matrix elements $\langle \Psi^{(a)} | P(1+A) | \Psi^{(b)} \rangle$) is equally trivial, under
the same condition.

\begin{acknowledgments}
The author thanks Leonid Gurvits for useful discussions and, in particular, for
bringing to the author's attention the ``permanent as an inner product'' construction.
\end{acknowledgments}

\end{document}